\documentstyle[11pt,paspconf]{article}

\begin{document}

\title{Resonance Paramagnetic Relaxation and Alignment of Ultrasmall Grains}

\author{A. Lazarian \altaffilmark{1}}
\affil{Princeton University Observatory, Princeton, NJ 08544}


\altaffiltext{1}{Present address: CITA, Univ. of Toronto, e-mail:
lazarian@cita.utoronto.ca} 



\begin{abstract}

A mechanism of enhanced paramagnetic relaxation for rapidly rotating 
interstellar grains is presented. 
We show that the Barnett magnetization that arises from grain rotation ensures 
that paramagnetic absorption happens at its maximum efficiency, i.e.
the conditions for paramagnetic resonance are automatically fulfilled.
The differences between the predictions of classical
Davis-Greenstein relaxation
and the process which we refer to as ``resonance relaxation'' are
most pronounced for grains rotating faster than 10~GHz.
Microwave polarization is likely to be an impediment for cosmic
microwave background studies, but can provide a good tool for
studying galactic magnetic field.

\end{abstract}




\section{Introduction}

In my other paper in this volume (Lazarian 1999) I discuss
the ``anomalous'' emissivity
and the two alternative explanations that this emission was
given in Draine \& Lazarian (1998a,b; 1999). Here I discuss
polarization that can arise from small rotating grains,
as I believe that this mechanism provides a more natural explanation
for the observed emissivity. The polarization of emission arising
from large magnetic grains
is discussed in Draine \& Lazarian (1999).

While paramagnetic alignment was one of the first mechanisms of
grain alignment to be discussed in the literature (Davis \& Greenstein 1951), 
it has been believed that ultrasmall
grains rotate too fast to be subjected to paramagnetic relaxation.
Indeed, it is easy to show that paramagnetic response of candidate
materials to oscillating magnetic field
is suppressed at $10^{10}-10^{11}$~Hz, which are typical
frequencies of ultrasmall grain rotation. As from the very beginning
of the paramagnetic research it has been taken for granted that 
paramagnetic relaxation does not depend on
whether a grain is rotating in magnetic
field or magnetic field is rotating about the grain the conclusion about
marginal alignment of rapidly rotating grains seemed self-evident. We show that
this assumption is incorrect as it disregards an important effect,
namely, the Barnett magnetization. 
We prove that, if this magnetization is accounted for, 
paramagnetic relaxation is dramatically enhanced at high frequencies. 

We call this effect ``resonance paramagnetic relaxation''
and discuss it's implications for the alignment of 
small carbonaceous and silicate grains. The results that this paper
deals with were first announced in Lazarian \& Draine (1998) and
will be discussed in detail in our forthcoming paper (Lazarian \& Draine
1999).

In what follows I compare Davis-Greenstein and resonance relaxation
processes (section~2) and discuss microwave polarization that is
expected from various ISM phases (section~3).

\section{Davis-Greenstein and Resonance Relaxation}

One of the mechanisms that can provide  alignment is the
paramagnetic dissipation mechanism suggested by Davis and
Greenstein half a century ago as a means of explaining the
polarization of starlight
(Davis \& Greenstein 1951). It's main idea is  
straightforward: for a spinning paramagnetic grain
the component of interstellar magnetic field
perpendicular to the grain angular velocity varies in grain coordinates 
resulting in time-dependent magnetization, associated 
energy dissipation, and a torque acting on the grain.
 As a result grains  tend to rotate
with angular momenta parallel to the 
interstellar magnetic field. As the the axis of maximal inertia 
of the rotating grain is aligned with the angular momentum
(see Purcell 1979)
the anisotropy in grain rotation
causes alignment with the long axis  grain perpendicular to the
interstellar magnetic field.

Grain should have unpaired electrons to be susceptible to paramagnetic
relaxation. Fortunately, even very small grains are likely to contain
paramagnetic species. Even in the absence of paramagnetic ions,
UV radiation, X-, $\gamma$-, and cosmic rays 
 create the whole gamut of different 
free radicals that are known to have paramagnetic properties.

However, it is known that the paramagnetic response drops
if the oscillating magnetic field has frequency $\omega>1/\tau_2$, where
$\tau_2$ is spin-spin relaxation time, which is
essentially the Larmor precession period of a spin in the
field of its neighbors. We expect $\tau^{-1}_2>10$~GHz.
Does this mean that the alignment is not efficient for
higher frequencies of grain rotation?

We claim that the traditional picture
of paramagnetic relaxation is incomplete (Lazarian \& Draine 1999). It
 disregards magnetization that develops due to grain rotation and
becomes increasingly 
important at high angular velocities.

It is well known
that a paramagnetic body
rotating in field-free space  
develops a magnetic moment due to the
Barnett effect (Landau \& Lifshitz 1960), which can be understood in
terms of body sharing part of its angular momentum with the spin system.
Therefore the implicit assumption in Davis \& Greenstein (1951)
that the dissipation within a {\it grain rotating} in a stationary 
magnetic field is equivalent to the dissipation within a {\it stationary
grain} in a rotating magnetic field is clearly not exact.  

Paramagnetic dissipation in a grain depends upon the imaginary 
part of the magnetic susceptibility $\chi^{''}$, which characterizes the
phase delay between grain magnetization and the rotating
magnetic field.

The time of paramagnetic relaxation is inversely proportional to
$K=\chi^{''}/\omega$ (see Davis \& Greenstein 1951). 
For {\it a stationary
grain} its magnetic response to the oscillation field
can be approximated (Draine \& Lazarian 1999):
\begin{equation}
K(\omega)\approx \frac{\chi_0}{(1+(\omega \tau)^2)^2}
\end{equation}

Calculations in Lazarian \& Draine (1999) provide a different expression
for $K(\omega)$, if the grain is rotating in a {\it stationary magnetic field}
and $\omega\gg \tau_2$:
\begin{equation}
K(\omega)\approx 0.5 \chi_0 \tau_2 
\frac {1}{1+\gamma^2 g_s^2 \tau_1 \tau_2
B^2}~~~,
\end{equation}
where $g_s$ is the gyromagnetic ratio $\approx 2$, $\gamma\equiv
e/(2 m_e c)\approx 1.76\times 10^6$~s$^{-1}$ gauss$^{-1}$, $B$
is the magnetic field intensity, $\tau_1$ and $\tau_2$ are spin-spin
and spin-lattice relaxation times. The latter time is essentially
the time over which the spin system can transfer its energy to the
lattice.
Estimates in Lazarian \& Draine (1999)
have shown that, for typical interstellar magnetic field, 
 the factor $\gamma^2 g_s^2 \tau_1 \tau_2B^2<1$ 
and therefore $K(\omega)\approx 
0.5 \chi_0 \tau_2$. 

As the magnetic alignment time $t_m$ is inversely proportional to $K(\omega)$
it is evident that for $\omega\gg \tau_2$ magnetic alignment time
predicted by DG theory, is much higher than that for resonance relaxation. Grain alignment
depends on the ratio $\delta\equiv t_d/t_m$, where $t_d$ is the
time scale of rotational damping; large $\delta$ correspond to good
alignment. Therefore it is evident that DG mechanism underestimates
the alignment of very rapidly rotating grains. 

It is instructive to consider why paramagnetic dissipation may proceed
effectively in a very rapidly rotating body while it may be ``saturated''
if the body is stationary and the magnetic field
rotates about it. A quantum
picture may be useful for the purpose: paramagnetic
absorption happens when the energy  of the electromagnetic quantum
is equal to the difference between energy levels within the spin
system. In the Davis-Greenstein picture the energy levels arise from
the interactions of adjacent spins and for a particular electron
the energy difference depends on whether its spin is directed along or
against  the magnetic field in its vicinity. This field intensity
is determined by the spin concentration and therefore for sufficiently
high frequencies the density of appropriate levels drops and so
does paramagnetic absorption. Barnett magnetization 
alters the picture
as the energy of an individual the spin becomes different when the spin
is directed parallel or anti-parallel to the
 angular velocity; the higher the rotational 
velocity, the greater the splitting. Therefore absorption of electromagnetic
quanta proceeds efficiently for high frequencies.

\section{Microwave Polarization from ISM Phases}

Emission from a rotating dipole is highly polarized. Therefore if
small grains are aligned, we expect to see polarization of the
microwave emission.

Alignment of tiny rotating grains depends on $\delta$ and consequently on
the damping time $t_d$. This time varies for 
various ISM phases (Draine \& Lazarian 1998b). Calculations in Lazarian \& Draine (1999) show that
the polarization of microwave emission from small grains
in cold ISM can reach $6\%$ in the range $10-30$~GHz. This may potentially 
present a  problem for cosmic microwave background experiments. In
molecular clouds the polarization can be as high as $10\%$ for the
same frequency range and may trace magnetic field deep inside
dark clouds\footnote{The issue of $K(\omega)$  saturation for
high values of magnetic field $B$ is unclear at the moment. It 
can be resolved on the basis of laboratory measurements of $\tau_1$ for
small grains.} where large grains are not aligned (Lazarian, Goodman \&
Myers 1997) and polarimetry at other wavelength fail.
In hot and warm ISM phases grain coupling with plasma is large
and $t_d$ is reduced. As a result, microwave polarization of a fraction
percent is achievable.

CMB measurements are intended at high galactic latitudes. For those
directions microwave foreground polarization mostly originates in
cold phase of the ISM. Overall polarization of the order of 3\% is
expected.

\section{Summary}

~~~~~1. The efficiency of paramagnetic relaxation depends on whether
a grain rotates in a stationary magnetic field or  magnetic field
rotates about a stationary grain. This difference arises from the Barnett
magnetization of a rotating grain. 
The original Davis-Greenstein theory 
disregarding this magnetization is incomplete, though sufficiently accurate
for slowly rotating grains. Grains rotating faster than $\sim 10$~GHz
experience resonance relaxation that enables alignment of very small
grains (e.g. $a<10^{-7}$~cm).

2. Alignment of small grains varies from one ISM phase to another
and so does the polarization of microwave emission. The expected
degree of polarization is of the order of a few percent in the range
$10-30$~GHz.

3. Polarization of microwave emission can become an important tool
for studies of magnetic field structure in molecular clouds.

\acknowledgments
I acknowledge discussions with 
Dick Crutcher, Chris McKee and Phil Myers. The research was supported
by NASA grants NAG5-2858, NAG5-7030, and CITA Senior Fellowship.


%
%

%

\end{document}